\documentstyle[12pt]{article}
\begin{document}

\title{Fermion zero modes in  Painlev\'e-Gullstrand black
hole.}
\author{G.E. Volovik\\
Low Temperature Laboratory, Helsinki University of
Technology\\
P.O.Box 2200, FIN-02015 HUT, Finland\\
and\\
L.D. Landau Institute for
Theoretical Physics\\  Kosygin Str. 2, 117940 Moscow, Russia
}
\maketitle
\begin{abstract}
{Painlev\'e-Gullstrand metric of the black hole allows to
discuss the fermion zero modes inside the hole. The
statistical mechanics of the fermionic microstates can be
responsible for the black hole thermodynamics. These fermion
zero modes also lead to quantization of the horizon area.
 }
\end{abstract}

\section{Introduction}

In general relativity there are different nonequivalent
metrics $g_{\mu\nu}$, which describe the same gravitational
object. Though they can be obtained from each other by
coordinate transformations, in the presence of the event
horizon they produce the nonequivalent quantum vacua.  Among
the other metrics used for description of the black hole, the
Painlev\'e-Gullstrand metric
\cite{Painleve} has many advantages \cite{Martel,Schuetzhold}
and now it acquires more popularity (see e.g.
\cite{ParikhWilczek};
on extension of Painlev\'e-Gullstrand metric to the
rotating black hole see
\cite{Doran}). Also such a stationary but not static metric
naturally arises in the condensed matter analogs of gravity
\cite{Unruh,Visser,PhysRepRev,JacobsonVolovik,Stone}. Here
using the Painlev\'e-Gullstrand metric we consider the
structure of the low energy fermionic microstates in the
interior of the black hole and their contribution to the
black hole thermodynamics.

The interval in the Painlev\'e-Gullstrand space-time is
\begin{equation}
 ds^2=-\left(c^2- v_{\rm s}^2(r)\right)dt^2+ 2v_{\rm
s}(r)drdt+dr^2+r^2d\Omega^2~.
\label{Painleve}
\end{equation}
In the acoustic black and white holes,
$v_{\rm s}(r)$ is the radial velocity of the fluid, which
produces the effective metric for acoustic waves -- phonons --
propagating in the liquid \cite{Unruh,Visser,PhysRepRev}.
For the gravitational field produced by the
point source of mass $M$, the function  $v_{\rm s}(r)$ has
the form
\begin{equation}
v_{\rm s}(r)=\pm c\sqrt{r_h\over r} ~,~r_h={2MG}~,
\label{VelocityField}
\end{equation}
where $r_h$ is the radius of the horizon; $G$ is the Newton
gravitational constant; $c=\hbar=1$. The
Painlev\'e-Gullstrand metric breaks the time reversal
symmetry: the time reversal operation transforms black hole
to the white hole (see also
\cite{Schuetzhold}). The minus sign in
Eq.(\ref{VelocityField}) gives the metric for the black hole.
In the fluid analog of gravity this corresponds to the
liquid flowing inward. The plus sign characterizes the white
hole and correspondingly the flow outward in the fluid
analogy. The time reversal operation reverses the direction
of flow.

The Painlev\'e-Gullstrand metric describes the space-time both
in exterior and interior regions. This space-time, though not
static, is stationary. That is why the energy in the interior
region is well determined.  Moreover, as distinct from the
Schwarzschild metric, the particle energy spectrum
$E({\bf p})$ (the solution of equation
$g^{\mu\nu}p_\mu p_\nu + m^2=0$ where $p_0=-E$) is well
defined  for any value of the momentum ${\bf p}$. This allows
us to determine the ground state (vacuum) of the Standard
Model in the interior region and the thermal states -- the
black hole matter. We consider here only the fermionic vacuum
of the Standard Model, assuming that the Standard Model is
an effective theory and thus the bosonic fields are the
collective modes of the fermionic vacuum.

\section{Fermi surface for Standard Model fermions inside
horizon}

We shall see that the main contribution to the thermodynamics
of the black hole comes from the very short wave length of
order of the Planck length. That is why the ultraviolet
cut-off must be introduced.  We introduce the cut-off
using the nonlinear dispersion of the particle spectrum in the
ultrarelativistic region, which violates
the Lorentz symmetry at short distances.
Such nonlinear dispersion of the particle spectrum is now
frequently used both in the black hole physics and cosmology
\cite{BHdispers,Corley,BHlaser,Starobinsky}. We shall use
the superluminal dispersion, for which the velocity of the
 particle becomes superluminal at very high momentum.
For the simplest superluminal dispersion the energy spectrum
of the fermionic particle in the  Painlev\'e-Gullstrand metric
becomes
\begin{equation}
 E=p_r v_{\rm s}\pm c\sqrt{ p^2 +{p^4\over p_0^2}}
\label{SuperluminalSpectrum}
\end{equation}
where   $p_r$ is the radial
momentum of the particle; $p_0$ is plays the role of
the cut-off momentum which is somewhat less than the Planck
momentum
$p_{\rm Planck}$; and we neglected all the masses of the
Standard Model fermions, since they are much less than the
charactersitic energy scales.

Because of the possibility of the superluminal
propagation, the surface $r=r_h$ is not the true
horizon. This surface marks the boundary of the ergoregion:
at $r<r_h$ particles with positive square root in
Eq.(\ref{SuperluminalSpectrum})  can acquire negative
energy. As a result, at $r<r_h$ the Fermi surface appears --
the surface in the 3D momentum space, where the energy of
particles is zero,
$E({\bf p})=0$. For the spectrum in
Eq.(\ref{SuperluminalSpectrum}) the surface is given by
equation, which expresses the radial momentum in terms of the
transverse momentum $p_\perp$:
\begin{equation}
  p_r^2(p_\perp)= {1\over 2}p_0^2 (v_{\rm s}^2-1) -p_\perp^2
\pm \sqrt{{1\over 4}p_0^4 (v_{\rm s}^2-1)^2 -p_0^2 v_{\rm
s}^2p_\perp^2}
\label{FermiSurface}
\end{equation}
This surface exists at each point ${\bf r}$ within the
horizon (ergosurface), where
$v_{\rm s}^2>1$. It exists only in the restricted range of
the  transverse momenta, with the restriction provided by the
cut-off parameter $p_0$:
\begin{equation}
   p_\perp<{1\over 2} p_0\left|v_{\rm s} -{1\over
v_{\rm s}}\right|~.
\label{MomentumRestriction}
\end{equation}
This means that the Fermi surface is a closed surface in the
3D momentum space ${\bf p}$.

The Fermi surface provides the finite density of
fermionic states (DOS) at $E=0$
\begin{eqnarray}
N(E=0)=N_F\sum_{{\bf p},{\bf r}} \delta (E({\bf p}))=
\label{DOSdefinition1}\\
{4\pi
N_F\over (2\pi)^3}
 \int_0^{r_h}r^2dr\int d^3p~
\delta\left(p_r v_{\rm s}+ c\sqrt{ p^2 +{p^4\over p_0^2}}
\right)=
\label{DOSdefinition2}\\
 {  N_F\over  \pi }
 \int_{0}^{r_h}r^2dr\int_0^{p_\perp^2(r)}
{d(p_\perp^2)\over |v_G|}~.
\label{DOSdefinition3}
\end{eqnarray}
Here $N_F=16N_g$  is the number of the massless
chiral fermionic species  in the Standard Model with $N_g$
generations; $v_G$ is the radial component of the group
velocity of particles at the Fermi surface:
\begin{equation}
  v_G(E=0)={dE\over dp_r}= \mp \sqrt{
\left(v_{\rm s} -{1\over v_{\rm s}}\right)^2 -4
{p_\perp^2\over p_0^2}} ~.
\label{GroupVelocity}
\end{equation}
Integration over $p_\perp^2$ in Eq.(\ref{DOSdefinition3})
gives for the density of states
\begin{equation}
N(E=0)= {  N_Fp_0^2\over  \pi }
 \int_{0}^{r_h}r^2dr
  \left|v_{\rm s}-{1\over v_{\rm s}}\right|
={4N_F\over  35\pi }p_0^2r_h^3 .
\label{DOSClassical}
\end{equation}
The main contribution to DOS and thus to the
thermodynamics comes from the momenta $p$ comparable  with
the cut-off momentum $p_0$. That is why all the masses of
fermions were neglected.

The DOS $N(E=0)$ determines the thermodynamics of the black
hole matter at $T\neq 0$. The thermal energy ${\cal E}(T)$ and
entropy ${\cal S}(T)$ carried by the Standard Model fermions
in the interior of the black hole with nonzero temperature is
\begin{equation}
{\cal E}(T)= N(0)\int dE E f(E/T)  ={\pi^2\over
6}   N(0) T^2~,~{\cal S}(T)=    {\pi^2\over 3}
N(0)T ~,
\label{ThermalEnergyEntropy}
\end{equation}
where $f(x)=1/(e^x+1)$ is the Fermi distribution function.

Here we considered the large temperature as compared to the
Hawking temperature $T_{\rm H}=\hbar c/4\pi r_h$
\cite{HawkingNature}. At lower energies the discrete nature
of the spectrum of the Standard Model fermions bound to the
black hole becomes important.

At $T\sim T_{\rm H}$ the entropy becomes of order
 \begin{equation}
{\cal S}(T\sim T_{\rm H})\sim N_Fp_0^2r_h^2
\label{ThermalEntropyAtHawking}
\end{equation}
The cutoff momentum $p_0$ can be expressed in terms of
the effective gravitational constant $G$, which is
determined by the same cutoff according to the Sakharov
effective gravity \cite{Sakharov}. The effective
gravity is obtained by integration over the vacuum fermions,
and thus all fermionic species must add to produce the
inverse effective gravitational constant: $G^{-1} \sim p_0^2
N_F$. Thus the thermal entropy
in Eq.(\ref{ThermalEntropyAtHawking}) is scaled as
$G^{-1}$, i.e. ${\cal S}(T\sim T_{\rm H})\sim  r_h^2/G$.
The same occurs with the Bekenstein-Hawking entropy
of the  black hole, $S_{\rm BH}=\pi r_h^2/G$. As it was first
shown by Jacobson,
$S_{\rm BH}$  is renormalized by the same quantum
fluctuations as the effective gravitational constant $G$, and
thus is proportional to  $G^{-1}$
\cite{JacobsonInduced}.)
Thus the thermal entropy of the Standard Model
fermionic microstates within the black hole at
$T\sim T_{\rm H}$ has the same behavior and the same order
of magnitude as the Bekenstein-Hawking entropy
of the black hole.

\section{Discrete energy levels inside horizon}

Now we proceed to the low energy, where the quantization is
important and gives discrete energy levels for the Standard
Model fermions within the horizon.  Since the momenta of
particles are large compared to the size of the horizon, one
can use the quasiclassical approximation for the radial
motion and the Bohr-Sommerfeld quantization rule. We consider
here the low energy states whose energy $E$ is much less than
the characteristic energy scale of the Fermi liquid:  $E \ll
p_0c$. In this limit the classical trajectories, which
determine the Bohr-Sommerfeld quantization, can be
obtained by perturbation theory. Let us start with the zero
order trajectories, i.e. trajectories with $E=0$.  After
quantization of the azimuthal motion, one obtains the
following dependence of the radial momentum
$p_r$ on $r$, which determines the classical trajectories
along the radius at a given value of the angular momentum
$L$ (compare this with Eq.(\ref{FermiSurface})):
\begin{equation}
  p_r^2(r,E=0,L)= {1\over 2}p_0^2 (v_{\rm s}^2-1) -{L^2\over
r^2}
\pm \sqrt{{1\over 4}p_0^4 (v_{\rm s}^2-1)^2 -p_0^2 v_{\rm
s}^2{L^2\over r^2}}~.
\label{SuperluminalTrajectoryZeroE}
\end{equation}
Since for typical bound states one has
$L\gg 1$, the difference between expressions $L(L+1)$,
$(L+{1\over 2})^2$ and $L^2$ is not important.

The trajectories in Eq.(\ref{SuperluminalTrajectoryZeroE}) are
closed: there are two turning points on each trajectory.
Particle moves back and forth between the zeroes
$r_1$ and
$r_2$ of the square root in the right hand side of
Eq.(\ref{SuperluminalTrajectoryZeroE}):
\begin{equation}
 \left|v_{\rm s}(r_{1,2})-{1\over v_{\rm
s}(r_{1,2})}\right| = {2L\over  p_0 r_{1,2}}~.
\label{TurningPoints}
\end{equation}
At the turning points the group velocity of the particle $v_G$
in Eq.(\ref{GroupVelocity}) becomes zero and changes sign to
the opposite. This is very similar to Andreev reflection
\cite{Andreev}: the velocity changes sign
after reflection while the momentum
$p_r$ does not.

For  nonzero but small energy, $E\ll p_0c$,  the
trajectories are obtained by perturbation theory. The first
order correction gives
\begin{equation}
 p_r(r,E,L)= p_r(r,E=0,L) + {dp_r\over dE}E=p_r(r,E=0,L) +
{E\over v_G(E=0)} ~.
\label{SuperluminalTrajectoryNonZeroE}
\end{equation}
The Bohr-Sommerfeld quantization gives
\begin{equation}
 2\pi ( n_r+\gamma(L))=\oint drp_r(r,E=0,L)+E\oint {dr\over
v_G}~,
\label{Quantization}
\end{equation}
where $n_r$ is the radial quantum number; and $\gamma(L)$ is
the parameter of order unity, which is not determined within
this quantization scheme.
Numerical integration of  $\oint drp_r(r,E=0,L)$ shows that
it is very close to the following equation
(\cite{Eltsov})
\begin{equation}
\oint drp_r(r,E=0,L)= \pm {
3\sqrt{3}\pi\over 2}
\left(L_{max}-L\right)~,~L_{max}={p_0r_h\over  3\sqrt{3}}.
\label{Numerics}
\end{equation}
As a result one obtains the following
equidistant energy levels for each $L$:
\begin{eqnarray}
E=\omega_0(L) \left( n_r +\gamma(L) \mp {
3\sqrt{3} \over 4}
\left(L_{max}-L\right)\right)~,
\label{EnergyLevels}
\\{\pi\over \omega_0(L)}=
  \int_{r_1}^{r_2} dr{1 \over \sqrt{ \left(v_{\rm s} -{1\over
v_{\rm s}}\right)^2 -4  {L^2\over r^2p_0^2}}}~.
\label{InterlevelDistance}
\end{eqnarray}
 In the two limiting cases the interlevel distance
$\omega_0(L)$ has the following values:
\begin{equation}
 \omega_0(L\ll p_0r_h)\approx
{\pi\hbar c\over r_h \ln{p_0r_h\over L}} =
T_{\rm H}{4\pi^2\over
\ln{p_0r_h\over L}}~,~~ \omega_0(L\rightarrow L_{max})=
{3\hbar c\over r_h }=12\pi T_{\rm H}~.
\label{EnergyLevelsLowL}
\end{equation}

Thus for each spherical harmonics $L,L_z$ there are bound
states in the black hole interior whose energy as a function
of the radial quantum number $n_r$ crosses zero energy level
at  $n_r\approx \pm { 3\sqrt{3}\over
4}\left(L_{max}-L\right)$. These are the branches of the
fermion zero modes. The total number of such branches is
\begin{equation}
N_{\rm zm}=2N_F\sum_{L<L_{max}}(2L+1)\approx 2 N_FL_{max}^2=
{2\over 27}N_Fp_0^2 r_h^2 \sim {A\over G}~,
\label{NumberZeroModes}
\end{equation}
where $A=4\pi r_h^2$ is the area of the black hole horizon.
The estimation of the density of states remains the same as
in Eq.(\ref{DOSClassical}) which was obtained within
the Fermi-surface approach:
\begin{eqnarray}
N(0)=2N_F\sum_L   {2L+1\over \omega_0(L)}=2N_F\int_0^\infty
d(L^2)  {1\over \omega_0(L)}=\\
={ N_Fp_0^2\over  \pi }
 \int_{0}^{r_h}r^2dr ~
  \left|v_{\rm s}-{1\over v_{\rm s}}\right|
=  {4N_F\over  35\pi }p_0^2r_h^3 ~.
\label{DOSNew}
\end{eqnarray}

\section{Fermion zero modes: vortex vs black hole}

The energy spectrum of the ultrarelativistic fermions within
the black hole in Eq.(\ref{EnergyLevels}) resembles the
spectrum of fermionic bound state within the core of vortices
in Fermi superfluids and superconductors:  The energy levels
are also equidistant there \cite{Caroli}.  The energy
spectrum of fermion zero modes in the vortex core depends on
two quantum numbers appropriate for the states within the
linear oblect, linear and angular momenta along the vortex
axis:
\begin{equation}
E(L_z,p_z)=
\omega_0(p_z)(L_z +\gamma)~,
\label{VortexSpectrum}
\end{equation}
where parameter $\gamma$ is either
$0$ or $1/2$ depending on the type of the vortex  (see review
paper \cite{PhysRepRev}).  For
vortices with
$\gamma=0$  the energy levels with $L_z=0$ have exactly
zero energy. For such a vortex the entropy is nonzero
even at $T=0$. Each of the states with $E=0$ can be
either free or occupied by fermion. This gives the zero
temperature entropy
$\ln 2$ per each state $E=0$ with given $p_z$;
and thus the total entropy of the vortex at $T=0$ is
proportional to the length
$l$ of the vortex line:
\begin{equation}
{\cal S}(T=0)=N_{\rm zm}\ln 2~,~N_{\rm zm}\propto p_0l ~.
\label{EntropyZeroTVortex}
\end{equation}
Here $p_0=p_F$ is the Fermi momentum of the Fermi superfluid
and superconductors; it plays the role of the cut-off
parameter. It is interesting that, as in the case of the black
hole, the wavelength of fermions comprising the fermion zero
modes in the vortex core is much less than the core size. This
allows us to use the quasiclassical theory. However, even
within the quasiclassical theory one can, using the symmetry
or other arguments, find the value of the phase shift
$\gamma$ in the Bohr-Sommerfeld quantization scheme
\cite{PhysRepRev}.  That is why one can predict for which
vortex the system of the equidistant levels of fermions
contains the states with exactly zero energy.

For the fermionic states bound to the black hole the
parameter $\gamma(L)$ in Eq.(\ref{EnergyLevels}) is still
unknown. That is why one cannot say whether the system of  the
equidistant levels contains the level with zero
energy or not. If yes, then each state with zero energy
contributes the entropy $\ln 2$; and the total entropy
provided by the fermion zero mode at
$T=0$  is
\begin{equation}
{\cal S}(T=0)=N_{\rm zm}\ln 2 ~.
 \label{EntropyT0}
\end{equation}

\section{Discussion}

From  Eq.(\ref{NumberZeroModes}) it follows that for
the Painlev\'e-Gullstrand black hole the area of the black
hole horizon is expressed in terms of the integer valued
quantity:
\begin{equation}
Ap_{\rm Planck}^2=\sigma {\cal N}~,
\label{AreaQuantization}
\end{equation}
where  ${\cal N}$ is the number of
fermion zero modes within the black hole:  ${\cal N}=N_{\rm
zm}$; and $\sigma$ is of order unity.  This formula with
different values of the parameter $\sigma$ was
discussed in many modern theories of black holes (see
\cite{BekensteinQuantum,Kastrup} and references therein). It
was interpreted as quantization of the horizon area, with
${\cal N}$ being the quantum number which characterizes the
black hole as an `atom' \cite{BekensteinQuantum}. If one uses
$\sigma=4\ln 2$ as in  Ref. \cite{Mukhanov}, one obtains
the Bekenstein-Hawking entropy in Eq.(\ref{EntropyT0}).
${\cal N}$ was also interpreted as the number of
`constituents' of the black hole interior -- the
`gravitational atoms' \cite{Mazur}. In our case both
interpretations are applicable, though with some reservation.

The quantization of area in Eq.(\ref{AreaQuantization})
usually suggests that the spacing between the levels is
uniform  and is on the order of $dM/d{\cal N} \propto E_{\rm
Planck}^2/M$
\cite{BekensteinQuantum}. This is in agreement with the
Eq.(\ref{EnergyLevelsLowL}) for the interlevel distance
$\omega_0$ in the fermionic spectrum. However, the
quantization in terms of the number of fermion zero modes
suggests another possible interpretation:

The area $A$ of the black hole is a continuous parameter. When
it changes, the number of fermion zero modes $N_{\rm zm}(A)$
as a function of $A$ changes in step-wise manner at some
critical values of $A$. This is what happens, say, in the
interger quantum Hall effect, where the integer topological
charge
${\cal N}$  of the qauntum vacuum as a function of external
parameters has plateaus. If the external parameter is the
magnetic field $B$, then ${\cal N}(B)$ and the Hall
conductivity $\sigma_{xy}(B)=(e^2/h){\cal N}(B)$ change
abruptly when the critical values of the magnetic field $B$
are crossed. Similar behavior of the topological charge
${\cal N}$ of the quantum vacuum occurs in other quasi-2D
fermionic systems too. For example, the momentum-space
topological invariant
${\cal N}$ of the film of quantum
fermionic liquids is a step-wise function of the continuous
parameter -- the thickness of the film  (see Chapter 9 in
Ref. \cite{Exotic}).

${\cal N}$ in Eq.(\ref{AreaQuantization})
can be also related to the number of `constituents' as
suggested in Ref. \cite{Mazur}.  According to
the Fermi liquid description,
the number of thermal fermions
in the Fermi liquid at temperature $T$ is
$N_{thermal}\sim N(0)T \sim A/G$. According
to Eq.(\ref{ThermalEnergyEntropy}) each fermion carries energy
of order $T$. At
$T\sim T_{\rm H}$  their total thermal energy is on the order
of the mass
$M$ of the black hole, and they  carry the thermal entropy of
the order of the Hawking-Bekenstein entropy $S_{\rm BH}$.
Assuming that the whole mass $M$ of the black hole comes from
the thermal fermions within the horizon, one has
$M=E+3pV$, where $V=(4\pi/3)r_h^3$ is the volume within the
horizon and $p$ the pressure of the fermionic system. Using
the equation of state of thermal fermions forming the Fermi
surface,
$E=ST/2 =pV$,  which follows
from Eq.(\ref{ThermalEnergyEntropy}), one obtains the correct
relation between the mass, Hawking temperature and
Bekenstein-Hawking entropy of the black hole:
\begin{equation}
M=E+3pV=4E=2TS~.
\label{ThermodynamicRelation}
\end{equation}
The Bekenstein-Hawking entropy, $S=\pi r_h^2$, and the Hawking
temperature, $T_{\rm H}=1/4\pi r_h$, are reproduced for the
following relation between the cut-off parameter
$p_0$ and the Newton constant:
$G^{-1}=N_Fp_0^2/105\pi$.
Thus the  black hole fermionic matter at the Hawking
temperature can provide the mass and the entropy of the black
hole, and thus the thermal fermions excited within the
horizon can serve as `constituents'.

These constituents do not actually represent the
`gravitational atoms' which form the quantum vacuum and give
rise to the phenomenon of gravitation according to Ref.
\cite{Mazur2}. These are conventional elementary particles
of the Standard Model (quarks and leptons) who are excited
within the black hole. Their contribution is essential even
at the temperature as low as $T_{\rm H}$, because of the huge
density of fermionic states within the black hole.

On the other hand, these constituents have little to do with
the matter absorbed by the black hole during its formation.
The fermions, which form the matter within the black holes,
are all the fermions of the Standard Model, quarks
and leptons, which are highly ultrarelativistic. The black
hole metric emerging after collapse perturbs significantly the
spectrum of Standard Model fermions, so that the
Fermi surface appears which provides a huge density of states
at zero energy.  Essential part of these fermions have
momenta of order of Planck scale; for them the effective
gravitational theory probably is not applicable. In this
sense these fermions are close to the `gravitational atoms'
of the trans-Planckian physics.

We considered the vacuum of the Standard Model fermions and
their thermal states as viewed in the Painlev\'e-Gullstrand
metric. This vacuum is  substantially different from the
vacuum state as viewed by comoving observer. The
reconstruction of the vacuum within the black hole involves
the Planck energy scale,  and results depend on the cut-off
procedure. The cut-off procedure, on the other hand, depends
on the coordinate system used and it assumes the existence of
the preferred coordinate frame at high energy. That is why the
vacuum structure depends on the coordinate system.

Since the Planck energy scale is involved, it is not clear
whether the traditional description of the black hole
is applicable. Moreover, the stability of this new
vacuum is not guaranteed. In most of those condensed matter
systems, where the analog of the event horizon is possible,
the vacuum becomes unstable in the presence of horizon, i.e.
the quantum vacuum of the condensed matter resists to the
formation of a horizon \cite{PhysRepRev,Visser2}. Also the
huge density of states may generate the symmetry
breaking in the black hole interior, as it happens in the
core of vortices \cite{MakhlinVolovik} and cosmic strings
\cite{Naculich}.

Even if the black hole survives under such reconstruction
of the Standard Model vacuum, there is another problem to be
solved. When the thermal states of the fermionic black-hole
matter are considered, their energy and pressure must serve as
a source of gravitational field according to (maybe somewhat
modified) Einstein equations. This will certainly change the
field $v_{\rm s}$ which enters the black hole metric and thus
the energy spectrum will be modified, but will remain
equidistant for each $L$.

In conclusion, we considered the statistical mechanics of
fermionic microstates -- the Standard Model fermions -- in the
interior of the black hole. The fermion zero modes give the
correct dependence of the entropy of the
Painlev\'e-Gullstrand black hole on the area of the black
hole, on the number of fermionic species, and on the Planck
cut-off parameter. They also lead to quantization of the
horizon area. That is why the fermion zero modes can be the
true microstates, that are responsible for the thermodynamics
of the black hole.

I thank T. Jacobson, P. Mazur, R. Sch\"utzhold, A.
Starobinsky, W. Unruh and M. Visser for illuminating
discussions and criticism. This work  was supported in part
by the Russian Foundation for Fundamental Research and by
European Science Foundation.

\end{document}